\documentclass[ floatfix,twocolumn]{revtex4}
\usepackage{amsmath}
\usepackage{color}
\usepackage{amssymb}
\usepackage{graphicx}
\usepackage[utf8]{inputenc}
\usepackage{textcomp}
\usepackage{color}
\usepackage{esint}

\begin{document}

\title{ 
Altermagnets versus  Antiferromagnets}

\author{Vladimir P.Mineev}
\affiliation{Landau Institute for Theoretical Physics, 142432 Chernogolovka, Russia}

\begin{abstract}

Altermagnets are substances  with a momentum-dependent spin splitting of electron bands due to a specific crystal structure that is invariant under time reversal only in combination with rotations and reflections. The developed phenomenological approach allows  to obtain a spectrum of electron bands in  altermagnets corresponding to 
their magnetic group symmetry.
We consider antiferromagnetic, weak ferromagnetic, and ferrimagnetic structures. The calculated Berry curvature of electron bands in the altermagnet with zero spontaneous magnetisation occurs an odd function of momentum components whereas it is even in a weak ferromagnetic state. The Berry curvature in ferrimagnetic state consists of sum of even and odd terms.
\end{abstract}
\date{\today}
\maketitle

\section{Introduction}

The momentum-dependent spin splitting of electron bands in metallic collinear antiferromagnets in the absence of spin–orbit coupling was investigated first in several theoretical papers \cite{Noda2016,Okugawa2018,Ahn2019,Hayami2019}. 
Soon after, the term "altermagnets" \cite{Smejkal2022} was proposed for materials of this type. 
 Similar to antiferromagnets altermagnets are characterised by a vanishing net magnetisation. But unlike antiferromagnets, opposite spin configurations in an altermagnet are not related by translation or inversion, but are connected by  rotation and reflection symmetry of  crystal.
An approach to the systematic classification and description of magnetic materials was developed
\cite{Smejkal2022} based on
the spin groups formalism. This approach, implemented within the so-called exchange approximation, allows  to determine  momentum dependent spin splitting of electron bands.

The impact of the crystalline environment, enabled by the spin-orbit coupling, on the magnetic and electronic properties of an altermagnet  has been investigated in
the paper \cite{Fernandes2024} based on formalism of magnetic groups.The possible electron bands spin-splitting has been formally classified  according with time-reversal-odd, 
space-inversion-even  irreducible representations of centrosymmetric point groups. 
The concrete symmetry of magnetic state  originating from  paramagnetic state  of a particular substance was not considered.

Another approach to the metals with momentum dependent band splitting based on traditional symmetry classification of magnetic materials \cite{LL1984} which includes both nonrelativistic and relativistic  interactions  has been developed in \cite{MineevUFN}.  A detailed   application of this approach to  altermagnet CrSb has been recently reported in \cite{Mineev2026}.

 An altermagnetic state arises from the corresponding paramagnetic state in crystals containing multiple magnetic atoms in the crystallographic unit cell.
The symmetry point group of the paramagnetic state, $G_p=G\times R$, includes the operations of the crystal point symmetry group and time reversal $R$, independent of each other.
In the case of a non-symmorphic symmetry group, $G$ includes rotations and reflections accompanied by shifts on the half-period of the unit cell. An altermagnetic state arising  from the paramagnetic state is invariant under time reversal only in combinations with rotations and reflections.
An altermagnet  symmetry point group $G_a$ is a subgroup of $G_p$. From a pure symmetry perspective, there is no distinction between 
an antiferromagnetic state,
characterised by the appearance of finite magnetic moments of magnetic atoms and invariant in respect of operations of group $G_a$ in coordinate space,
and an altermagnetic state, characterised by the appearance of a momentum-dependent spin splitting of the electron bands  invariant in respect of operations of group $G_a$ in reciprocal space. Moreover, any antiferromagnet,  no matter metal or insulator,  with symmetry pointed above is at the same time an altermagnet with the same symmetry.

The ordered structure in the antiferromagnetic state can be determined by minimising of the Landau free energy expansion invariant under
all operations of the point symmetry group of the paramagnetic state
in powers of the order parameter given by linear combinations of the magnetic moments of the magnetic atoms ${\bf m}_i$ in the unit cell \cite{Dzyal1957,Kzial1957}.
For example, in the case of two magnetic atoms in a unit cell, these combinations have the form ${\bf M}={\bf m}_1+{\bf m}_2$ and ${\bf L}={\bf m}_1-{\bf m}_2$.
The ordering in the altermagnetic state does not have a similar visual representation.
However, each antiferromagnetic crystal structure
invariant under time reversal only in combination with rotations and reflections corresponds to an altermagnetic crystal structure with the same electron band symmetry.
Hence, just looking on ordering of magnetic moments in coordinate space of concrete antiferromagnet one can establish the momentum dependent spin splitting of electron bands in an altermagnet with the same symmetry.

Here, we discuss the parallel between antiferromagnetic and altermagnetic states, using as an example the possible antiferromagnetic structures of fluorides of transition metals with symmetry ${\bf D}_{4h}$.   These substances possess the spin-splitting of electron bands  typical for altermagnets. The magnon bands splitting in  MnF$_2$  and FeF$_2$ determined by altermagnet symmetry has been recently revealed by polarised neutrons scattering \cite{Faure2025,Sears2026}.

I.E.Dzyaloshinsky \cite{Kzial1957} investigated possible antiferromagnetic structures in paramagnets with symmetry ${\bf D}_{4h}$.
They are: collinear antiferromagnetic without spontaneous magnetisation, weak ferromagnetic antiferromagnetic and ferrimagnetic states. All these structures possess momentum dependent spin splitting of  electron bands. So, all these structures have property typical for altermagnets. In the present paper the electron energy spectra of the altermagnetic states with corresponding symmetry are described. The comparison of  electron band spin splitting derived in frame of spin symmetry groups approach and magnetic groups approach is presented.
Then the Berry curvature for each specific state is calculated.

 \section{Crystal structures and Band energies}
 
 \begin{figure}
\includegraphics
[height=.2\textheight]
{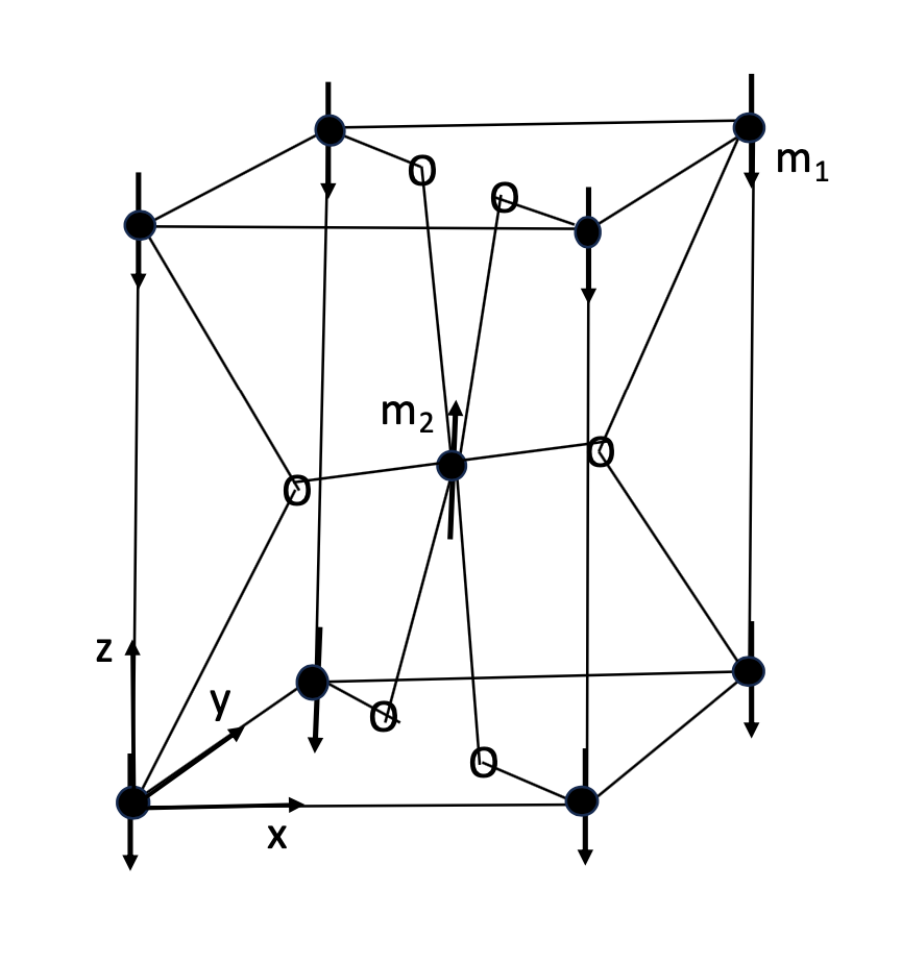}
 \caption{
Spin configuration for the case I. Solid circles indicate the positions of magnetic ions. The small circles correspond to  sites of nonmagnetic ions.
The antiferromagnetic ordering does not distort the tetragonal structure of paramagnetic state. }
\end{figure}

The crystals of fluorides of the transition metals   have tetragonal structure shown on Fig.1 and the symmetry group of its paramagnetic state is non-symmorphic point group 
\begin{eqnarray}
G_p={\bf D}_{4h}\times R=(E,C_{2}, 2C_{4}t,2U_2t,2U'_2)\times I\times R.
\label{sym}
\end{eqnarray}
Here, $C_2$ denotes a rotation on angle $\pi$ around [001] axis,  $2U_2t$ - corresponding rotations  around [100] or [010] axis accompanied by shift on half period $t=(a,a,c)/2$ along the prism diagonal, $2C_{4}t$ are rotations on angle $\pi/2$ around [001] axis ccompanied by shift $t$,
$2U'_2$ -rotations on angle $\pi$ around [110] or $[1\bar 10]$ axis, I is the operation of space inversion and $R$ is the operation of time inversion. 
I.E.Dzyaloshinskii \cite{Kzial1957} has shown that these crystals can exist in three magnetic states: antiferromagnetic state I without total magnetisation, and states II$_1$
and II$_2$ with small spontaneous magnetic moment which are weak ferromagnetic and ferrimagnetic states. 

 In the state I shown on Fig.1 the magnetic moments  of the metallic ions in two corresponding points of elementary cell are equal and of opposite signs. They  are directed along $\pm\hat z$ direction. 
 The electron energy as a function of momentum and spin  of each band has the form
 \begin{equation}
\varepsilon_{\alpha\beta}({\bf k})=\varepsilon_{\bf k}\delta_{\alpha\beta}+\mbox{\boldmath$\gamma$}_{\bf k}\mbox{\boldmath$\sigma$}_{\alpha\beta}.
\label{45}
\end{equation}
Here, $\varepsilon_{\bf k}$ and  $\mbox{\boldmath$\gamma$}_{\bf k}$ are
translation invariant even functions of momentum 
 and
$\mbox{\boldmath$\sigma$}=(\sigma_x,\sigma_y,\sigma_z)$ are the Pauli matrices. The general shape of these function is determined by symmetry demands.

The electron energy  spectrum possesses momentum dependent spin splitting determined by function $\mbox{\boldmath$\gamma$}_{\bf k}$. To find its shape
according to the spin groups  approach \cite{Smejkal2022} that is in non-relativistic exchange approximation  the symmetry of atoms positions and the magnetic moments directions  should be treated independently. Namely, its momentum dependence 
should correspond to one of even in respect of space inversion  irreducible representations  of the group of   symmetry of paramagnetic state, for instance $k_xk_y$. At the same time the spin vector $\mbox{\boldmath$\gamma$}_{\bf k}$ should be invariant in respect to
the symmetry operations acting only on arrows shown on Fig.1  which
keep the crystal  order. They  include operations of space inversion, any rotations around $\hat z$ axis $C\hat z=\hat z$, any rotations on angle $\pi$ around axis perpendicular to axis $\hat  z$
accompanied by time inversion $RU_2\hat z=\hat z$. Thus, we obtain 
\begin{eqnarray}
\mbox{\boldmath$\gamma$}_{\bf k}=
c\sin (k_xa)\sin(k_ya)\hat z.
\label{alt1}
\end{eqnarray}

On the contrary according to the magnetic groups theory  approach  the symmetry of function (\ref{45}) must coincide with symmetry of crystal structure in magnetic state which includes rotations and reflections acting both to atomic positions and to axial vectors of magnetic moments. Here, it is completely unimportant which microscopic mechanism - exchange or spin-orbit interaction provides the spin splitting of electron bands.
In particular
 the symmetry group of state I  consists of the following operations 
 \begin{eqnarray}
G_I==(E,C_{2}, 2C_{4}tR, 2U_2t,2U'_2R)\times I.
\label{symI}
\end{eqnarray}
The group $G_I$ is the subgroup of the group of symmetry of paramagnetic state Eq.(\ref{sym}). 
The vector  $\mbox{\boldmath$\gamma$}_{\bf k}$ possessing this symmetry that is relating to the unit representation of the group $G_I$  is
\begin{eqnarray}
\mbox{\boldmath$\gamma$}_{\bf k}=\gamma_1 \sin(k_zb) \left[\sin(k_ya)\hat x+\sin(k_xa)\hat y\right]\nonumber\\
+\gamma_2\sin (k_xa)\sin(k_ya)\hat z,
\label{alt1}
\end{eqnarray}
The equation (\ref{alt1}) defining the vector  $\mbox{\boldmath$\gamma$}_{\bf k}$ is the simplest possible expression that has the necessary symmetry properties.
The electron energy  spectrum found by means magnetic groups theory  approach possesses additional momentum dependent spin splitting in comparison with found 
in frame of spin group approach.

\begin{figure}
\includegraphics
[height=.2\textheight]
{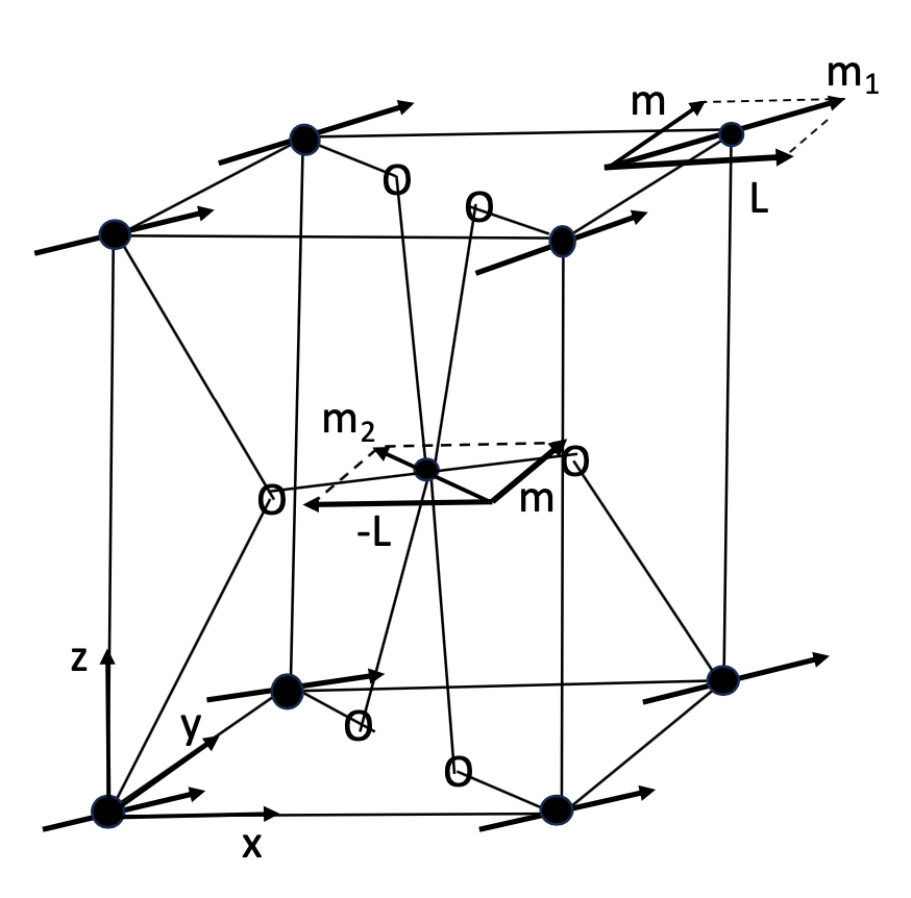}
 \caption{Spin configuration for the case II$_1$. In weak ferromagnet state the tetragonal symmetry of initial paramagnetic state is lowered  to orthorhombic one. 
 See the text.
}
\end{figure}

In the case II$_1$  considered in \cite{Kzial1957}
 the magnetic moments of the metallic ions  in two sub-lattices ${\bf m}_1$ and ${\bf m}_2$ 
  are located in plane (001) such that
 vector $({\bf m}_1-{\bf m}_2)/2={\bf L}$ is directed along [100] axis and vector $({\bf m}_1+{\bf m}_2)/2={\bf m}$ is directed along [010] axis. Thus, the state II$_1$   is an antiferromagnetic state possessing weak ferromagnetic moment.   See the Fig.2. Let us stress  that although
 the magnetic moment ${\bf m}$ shown on Fig2. looks like arising due to canting of sub-lattice magnetic moments ${\bf m}_1$ and ${\bf m}_2$,
 it is just effective moment induced by spin-orbit coupling taking into account in the Landau treatment developed in the paper  \cite{Kzial1957}. One can consider it as spontaneous magnetisation of the crystal lattice as whole. 
 Spontaneous magnetisation $2{\bf m}$
arises due to the presence of invariants like $A_{ij}m_iL_j$ allowed by symmetry \cite{Dzyal1957,Kzial1957}, which is ensured by  a spin-orbit
mechanism. Thus, the spin group treatment  to state II$_1$ is inapplicable.

 The symmetry group of   state II$_1$  is 
  \begin{eqnarray}
 G_{{II}_1}
= (E,U_{2y}t,C_2R,U_{2x}Rt)\times I,
 \label{2}
 \end{eqnarray}
 It should be noted that initial tetragonal symmetry of paramagnetic state is decreased to orthorhombic one in the state II$_1$.
 The energy of electron as a function of momentum in any substance with structure symmetric in respect of all the operations of this group has the 
form (\ref{45}) with vector
\begin{eqnarray}
\mbox{\boldmath$\gamma$}_{\bf k}=\gamma_1\sin( k_xa)\sin (k_yb)\hat x+gm[1+\alpha\sin^2(k_xa)]\hat y\nonumber\\+\gamma_2\sin(k_yb)\sin(k_zc)\hat z.
\label{7}
\end{eqnarray}
Here, $m=|{\bf m}|$.
The equation (\ref{7}) defining the vector  $\mbox{\boldmath$\gamma$}_{\bf k}$ is the simplest possible expression that has the necessary symmetry properties.

Weak ferromagnetism  has been investigated by I.V.Solovyev \cite{Solovyev1997}  based on the first principles band structure calculations. Recently the studies of this phenomenon have been continued in \cite{Naka2020,Naka2022} by means of the multi-orbital Hubbard model
and in \cite{,Solovyev2025} 
 with parameters derived from DFT (density functional theory).
It was noted  that "contrary to conventional wisdom, the spin canting is irrelevant to the Hall resistance, the Hall conductivity originates from the collinear component of antiferromagnetic order in the presence of spin-orbit coupling." This fact  is not so astonishing because the magnetic moment in antiferromagnetic weak ferromagnets is always proportional to the Neel vector components $m_i\propto A_{ij}L_j$ and its  interpretation by the "canting" of moments is naive one.

Let's now move on to the state II$_2$.
In the case II$_2$ (see \cite{Kzial1957}) the magnetic moments of the metallic ions in two corresponding poins of elementary cell  are not equal and directed parallel and antiparallel to [110] direction. In this case due to spin-orbit coupling the crystal acquires rhombic symmetry and losses non-symmorphic structure.
So, here we deal with ferrimagnetic state.
See Fig.3. The symmetry group of this state is 
\begin{eqnarray}
 G_{{II}_2}=(E,U'_{2xy},C_2R,U'_{2x\bar y}R)\times I.
 \label{3}
 \end{eqnarray}
 Corresponding vector $\mbox{\boldmath$\gamma$}_{\bf k}$ in spin part of the electron spectrum has the form
\begin{eqnarray}
\mbox{\boldmath$\gamma$}_{\bf k}=\gamma_1[\cos(k_xa)-\cos(k_ya)](\hat x-\hat y)\nonumber\\
+g(m_1-m_2)[1+\alpha\sin (k_xa)\sin( k_ya)](\hat x+\hat y)\nonumber\\
+\gamma_2\left [\sin( k_xa)+\sin (k_ya)\right ]\sin(k_zc)\hat z~~~
\label{9}
\end{eqnarray}
 invariant  in respect of all the operations enumerated in Eq.(\ref{3}).  
Here, $x$ and $y$ axis of coordinate system are chosen parallel to sides of rhomb in the base of parallelepiped shown of Fig.3.

\begin{figure}
\includegraphics
[height=.2\textheight]
{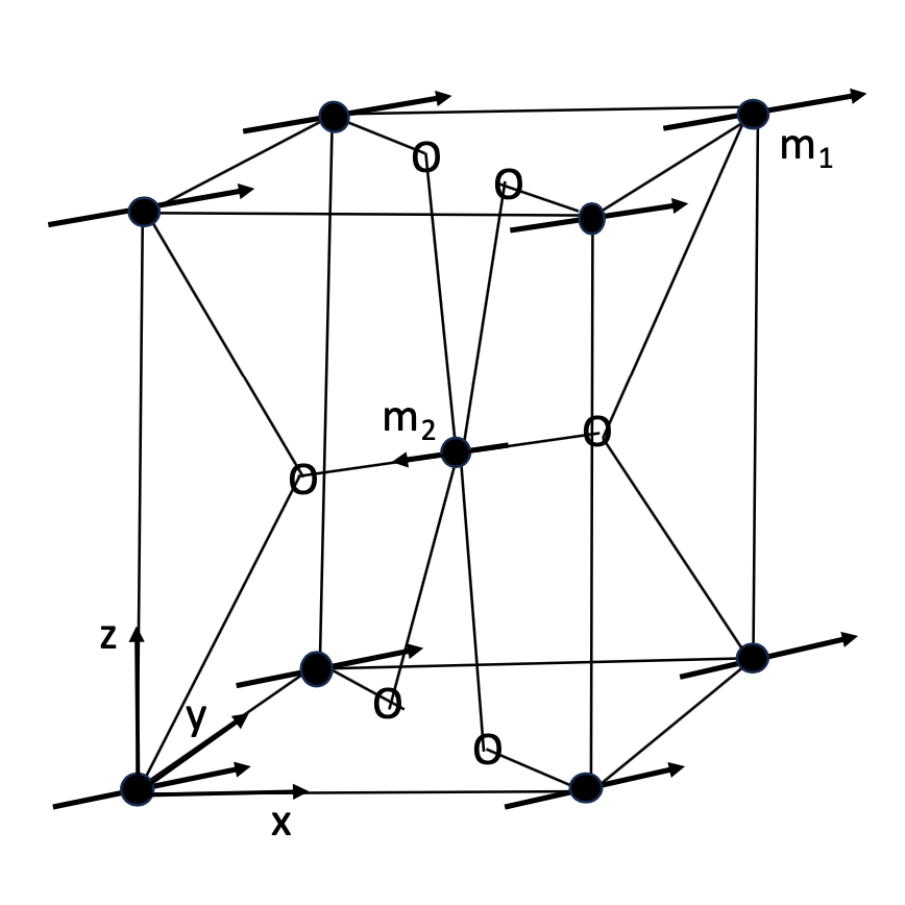}
 \caption{Spin configuration for the case II$_2$. In ferrimagnet state the crystal acquires symmorphic rhombic symmetry. See the text.}
\end{figure}

\section{Berry curvature}

The knowledge of band energy dispersions allows us to calculate the corresponding Berry curvature in the altermagnets with symmetry I, II$_1$ and II$_2$.

The eigenvalues of the matrix (\ref{45}) are
\begin{equation}
    \varepsilon_{\lambda}({\bf k})=\varepsilon_{\bf k}+\lambda\gamma,~~~~~~~~ \lambda=\pm,
\label{e3}
\end{equation}
where $\gamma=|\mbox{\boldmath$\gamma$}_{\bf k}|$.

The Berry curvature for each band  with $\lambda=\pm$ is \cite{Xiao2010} 
\begin{equation}
\Omega_{ij}^{\pm}({\bf k})=\mp e_{\alpha\beta\gamma}\frac{\gamma_\alpha}{2\gamma^3}\left (\frac{\partial\gamma_\beta}{\partial k_i}\frac{\partial\gamma_\gamma}{\partial k_j}  \right ).
\label{omega}
\end{equation}

Performing calculations   in the case I with vector $\mbox{\boldmath$\gamma$}_{\bf k}$  given by Eq. (\ref{alt1}) we obtain
\begin{eqnarray}
\Omega_{xy}^\pm
=\mp\frac{\gamma_1^2\gamma_2a^2}{8\gamma^3}
\sin(2k_xa)\sin(2k_ya)\sin^2(k_zb).
\label{ome}
\end{eqnarray}
We see that this expression  is the odd  function  of each of momentum components $k_x$ and $k_y$.
Similar property takes place for two other components of the Berry curvature tensor. 

Performing similar calculations for altermagnetic state II$_1$ with vector $\mbox{\boldmath$\gamma$}_{\bf k}$ given by Eq. (\ref{7}) we obtain
the   $xz$ component of the Berry tensor for this state 
\begin{eqnarray}
\Omega_{xz}^{\pm}=\mp \frac{\gamma_1\gamma_2gm~ac}{2\gamma^3}~~~~~~~~~~~\nonumber\\
\times\cos (k_xa)\cos (k_zc)\sin^2(k_yb)[\alpha\sin^2(k_xa)-1].
\end{eqnarray}
Now, the Berry curvature is  even function of momentum. 

To calculate the Berry curvature 
for the state II$_2$ by means of Eq.(\ref{omega}) one needs to rewrite Eq.(\ref{9}) in terms of orthogonal vectors $\hat X=\hat x-\hat y$, $\hat Y=\hat x+\hat y$, $\hat Z=\hat z$
and components of momentum $k_x,k_y$ in terms of components $k_X,k_Y$
\begin{eqnarray}
k_x=k_X\sin\varphi+k_Y\cos\varphi,\nonumber\\k_y=-k_X\sin\varphi+k_Y\cos\varphi.
\end{eqnarray}
Here, $\varphi$ is half of angle between the directions $\hat x$ and $\hat y$.
Then we obtain
\begin{eqnarray}
\Omega_{XZ}^{\pm}=\mp \frac{\gamma_1\gamma_2g(m_1-m_2)ac\sin\varphi}{2\gamma^3}\nonumber\\
\times \cos(k_zc)[ \sin(k_xa)+\sin(k_ya) ]\nonumber\\
\times\left\{[ \sin(k_xa)+\sin(k_ya) ](1+\alpha\sin(k_xa)\sin(k_ya))\right.\nonumber\\
+\left.\alpha[ \cos(k_xa)-\cos(k_ya) ]\cos(k_xa)\sin(k_ya)
\right\}.~~
\end{eqnarray}
The Berry curvature in ferrimagnetic state II$_2$ consists of sum of even and odd terms in respect of momentum components $k_X,k_Y$.

\section{Summary and discussion}

The developed phenomenological approach allows to find the electron band spectrum  in  crystal structures  invariant under time reversal only in combination with proper and improper rotations and reflections. 
The phenomenological approach enables calculation of the Berry curvature in a band structure with a given symmetry without cumbersome microscopic derivations within the framework of any specific model. It allows calculations to be performed using a simple single-band description of nonsymmorphic crystal symmetry.

There was considered three typical magnetic states.
They are:  state I without total magnetisation, and states II$_1$ and II$_2$ with small spontaneous magnetic moment arising due to spin-orbit interaction. 
The electron spectrum in the state I  possesses additional spin splitting in comparison with found by means spin group approach.

It was shown that in the altermagnet state without spontaneous magnetisation the Berry curvature is an odd function of momentum components.
 In our opinion this  is  the general property of altermagnetic state without spontaneous magnetisation what leads to the absence anomalous Hall effect in absence of magnetic field.

The Berry curvature, which  is even function of momentum, occurs  only in the weak ferromagnetic state I$_1$. In this respect the recent observation of magneto-optic Kerr effect in altermagnet hematit 
$\alpha$-Fe$_2$O$_3$ reported in \cite{Yoshimori2026} is not so astonishing. As it was shown already in the classic paper by I.E.Dsyaloshinskii \cite{Dzyal1957} this material 
is weak ferromagnetic antiferromagnet.  The reported recently the  Hall effect at zero external field
in antiferromagnet FeS  \cite{Takagi2025} has probably the same origin.

The expression for Berry curvature in the ferrimagnetic state contains both even and odd terms in momentum components.

The developed approach is 
general. In particular it is easy to apply it to any collinear altermagnet for instance to altermagnetic  analogues of antiferromagnetic MnC0$_3$ and CoCO$_3$ with rhombohedral crystal structure \cite{Dzyal1957}.

I am indebted to Banik Rai for numerous questions in particular related to possibility of existence spontaneous Hall effect in altermagnets without
bulk magnetisation.


\begin{thebibliography}{220}

\bibitem{Noda2016}Y. Noda, K. Ohno, and S. Nakamura, {\it Momentum-dependent band spin splitting in semiconducting MnO$_2$: A density functional calculation}, Phys. Chem. Chem. Phys. {\bf 18}, 13294 (2016).

\bibitem{Okugawa2018}T. Okugawa, K. Ohno, Y. Noda, and S. Nakamura, {\it Weakly spin-dependent band structures of antiferromagnetic perovskite LaMO$_3$ (M = Cr, Mn, Fe)}, J. Phys.: Condens. Matter {\bf 30}, 075502 (2018).

\bibitem{Ahn2019}K.-H.Ahn, A. Hariki, K.-W. Lee, and J. Kuneš, {\it Antiferromagnetism in RuO2 as d-wave Pomeranchuk instability}, Phys. Rev. B {\bf 99}, 184432 (2019).

\bibitem{Hayami2019} S.Hayami, Y.Yanagi, and H.Kusunose, {\it Momentum-Dependent Spin Splitting by Collinear Antiferromagnetic Ordering}, Journal of the Physical Society of Japan 88, 123702 (2019).

\bibitem{Smejkal2022} L.\v{S}mejkal,  J. Sinova, T. Jungwirth, Phys. Rev.X {\bf 12}, {\it Beyond Conventional Ferromagnetism and Antiferromagnetism: A Phase with Nonrelativistic Spin and Crystal Rotation Symmetry}, 031042 (2022).

\bibitem{Fernandes2024}R.M. Fernandes, V.S. de Carvalho, T. Birol, and R.G. Pereira,{\it Topological transition from nodal to nodeless Zeeman splitting in altermagnets}, Phys.Rev.B {\bf 109}, 024404 (20124).

\bibitem{LL1984} 
 L.D.Landau and E.M.Lifshitz, Course of Theoretical
Physics, Vol. 8: Electrodynamics of Continuous Media (Pergamon, Oxford, 1984). 


\bibitem{MineevUFN}V.P.Mineev, {\it Toroid, altermagnetic, and noncentrosymmetric ordering in metals},Uspekhi Fizicheskikh Nauk {\bf 195}, 1221 (2025)
[Physics - Uspekhi {\bf 68}, 1151 (2025)].

\bibitem{Mineev2026} V.P.Mineev, {\it Band splitting in the altermagnet CrSb}, Phys.Rev.B {\bf 114}, 065130 (2026), see corrections in arXiv:2602.15131.

\bibitem{Dzyal1957}I.E.Dzialoshinskii,  {\it Thermodynamic Theory of "Weak" Ferromagnetism In Antiferromagnetic Substances}, Zh. Exp. Theor. Fiz. {\bf 32}, 1547 (1957) [Soviet  Phys. JETP {\bf 5}, 1259 (1957)]; J. Phys. Chem. Solids {\bf 4},  241 (1958).
 
\bibitem{Kzial1957}I.E.Dzialoshinskii, {\it The magnetic structure of fluorides of the transition metals}, ZhETF(U.S.S.R.) {\bf 33}, 1454 (1957) [Soviet Physics JETP{\bf 6}, 1120 (1958)].

\bibitem{Faure2025} Q.Faure, D.Bounoua, V.Bale'dent, A.Gukasov,
V. O.Garlea, A.Ribeiro, J.G. Rau, S.Petit, P.McClarty, {\it Altermagnetism revealed by polarized neutrons in MnF$_2$}, arXiv: 2509.07087 [cond-mat.str-el].

\bibitem{Sears2026}J. Sears, V. O. Garlea, D. Lederman, J. M. Tranquada, and I. A. Zaliznyak, {\t Altermagnetic and dipolar splitting of magnons in FeF$_2$ },
arXiv: 2601.04303 [cond-mat.str-el].

\bibitem{Solovyev1997}I. V. Solovyev, {\it Magneto-optical effect in the weak ferromagnets LaMO$_3 ($M$=$Cr, Mn, Fe)}, Phys. Rev. B 55, 8060 (1997); 

\bibitem{Naka2020}M.Naka, S.Hayami, H.Kusunose, Y.Yanagi, Y. Motome, and H.Seo, {\it Anomalous Hall effect in $\kappa$-type organic antiferromagnets}, Phys. Rev. B {\bf 102}, 075112 (2020).

\bibitem{Naka2022}M.Naka, Y.Motome, and H.Seo, {\it Anomalous Hall effect in antiferromagnetic perovskites}, Phys. Rev. B {\bf 106}, 195149 (2022)

\bibitem{Solovyev2025}I. V. Solovyev, S. A. Nikolaev, and A. Tanaka, {\it Altermagnetism and Weak Ferromagnetism}, arXiv:2503.23735.

\bibitem{Xiao2010}Di Xiao, Ming-Che Chang, Qian Niu, {\it Berry phase effects on electronic properties}, Rev. Mod.Phys. {\bf 82}, 1959 {2010}.

\bibitem{Yoshimori2026} H.Yoshimochi, K. Yoshida, R. Oiwa, T. Nomoto, N.
D. Khanh, A. Kitaori, R. Takagi, R. Arita, S. Seki, {\it Large magneto-optical Kerr effect induced by collinear antiferromagnetic order}, arXiv:2601.13723.

\bibitem {Takagi2025} R.Takagi, R. Hirakida, Y. Settai1, R.Oiwa, H.Takagi, A.Kitaori, K.Yamauchi, H.Inoue, J. Yamaura, D.Nishio-Hamane, S.Itoh,
S.Aji, H.Saito, T.Nakajima, T.Nomoto, R.Arita, S.Seki, {\it Spontaneous Hall effect induced by collinear antiferromagnetic order at room temperature}, Nature Mat. {\bf 24}, 63 (2025).




\end{thebibliography}
\end{document}